\documentclass[10pt,conference]{IEEEtran}
\IEEEoverridecommandlockouts

\usepackage{cite}
\usepackage{amsmath,amssymb,amsfonts}
\usepackage{algorithmicx}
\usepackage{graphicx}
\usepackage{textcomp}
\usepackage[dvipsnames]{xcolor}
\usepackage{listings}
\usepackage[ruled,vlined]{algorithm2e}
\usepackage{subcaption}

\lstdefinestyle{mystyle}{
    basicstyle=\ttfamily\footnotesize,
    breakatwhitespace=false,         
    breaklines=true,                 
    captionpos=b,          
    keepspaces=true,                 
    numbers=left,                    
    numbersep=5pt,                  
    showspaces=false,                
    showstringspaces=false,
    showtabs=false,                  
    tabsize=2
}
\def\BibTeX{{\rm B\kern-.05em{\sc i\kern-.025em b}\kern-.08em
    T\kern-.1667em\lower.7ex\hbox{E}\kern-.125emX}}

\title{Chimera: A Hybrid Machine Learning Driven Multi-Objective Design Space Exploration Tool for FPGA High-Level Synthesis \thanks{This is an extended version of the conference paper published in the 22nd International Conference on Intelligent Data Engineering and Automated Learning (IDEAL 2021), which won the Best Paper Award. It is supported in part by the Xilinx Center of Excellence and Xilinx Adaptive Compute Clusters (XACC) program at the University of Illinois Urbana-Champaign.}}

\author{
    \IEEEauthorblockN{Mang Yu, Sitao Huang, Deming Chen}
    \\\IEEEauthorblockA{University of Illinois at Urbana-Champaign, \\Urbana IL 61820, USA
    \\{\{mangyu2,shuang91,dchen\}@illinois.edu}}
}

\begin{document}
\maketitle

\begin{abstract}
In recent years, hardware accelerators based on field-programmable gate arrays (FPGAs) have been widely adopted, thanks to FPGAs' extraordinary flexibility. However, with the high flexibility comes the difficulty in design and optimization. Conventionally, these accelerators are designed with low-level hardware descriptive languages, which means creating large designs with complex behavior is extremely difficult. Therefore, high-level synthesis (HLS) tools were created to simplify hardware designs for FPGAs. They enable the user to create hardware designs using high-level languages and provide various optimization directives to help to improve the performance of the synthesized hardware. However, applying these optimizations to achieve high performance is time-consuming and usually requires expert knowledge. To address this difficulty, we present an automated design space exploration tool for applying HLS optimization directives, called Chimera, which significantly reduces the human effort and expertise needed for creating high-performance HLS designs. It utilizes a novel multi-objective exploration method that seamlessly integrates active learning, evolutionary algorithm, and Thompson sampling, making it capable of finding a set of optimized designs on a Pareto curve with only a small number of design points evaluated during the exploration. In the experiments, in less than 24 hours, this hybrid method explored design points that have the same or superior performance compared to highly optimized hand-tuned designs created by expert HLS users from the Rosetta benchmark suite. In addition to discovering the extreme points, it also explores a Pareto frontier, where the elbow point can potentially save up to 26\% of Flip-Flop resource with negligibly higher latency.
\end{abstract}

\section{INTRODUCTION}
In recent years, hardware accelerators based on field-programmable grate arrays (FPGAs) have been widely adopted by both academia and industry. This is mainly because they offer performance close to customized hardware, as well as the flexibility of programmable devices. For this reason, a large number of FPGA-based accelerators have been created for tasks ranging from H.264 encoding to customized deep neural network inference \cite{liu2016high, zhou2018rosetta, zhang2017high, li2019implementing, zhao-bnn-fpga2017, han2017ese, jason1, Stereo}.

However, designing accelerators for complex applications on FPGAs requires an immense amount of human effort and expert knowledge, which hampers the high flexibility and short time-to-market granted by FPGA's programmability. Particularly, with the rise of fast-evolving machine learning algorithms, rapid design and deployment techniques for high-performance FPGA accelerators are very desirable.

In the light of these demands, the high-level synthesis (HLS) tools are developed to enable designers to describe hardware designs directly using high-level languages, which can significantly reduce the human efforts needed for creating customized accelerators \cite{HLS1, HLS2}.

However, the challenges of creating complex designs for FPGAs are not completely eliminated, as the HLS tools have various assumptions and limitations. So, compiling a software-oriented design in HLS without refactoring and applying optimization directives usually leads to poor accelerator performance. Therefore, as \cite{rupnow2011} pointed out, to achieve the best performance with HLS, the user needs advanced knowledge of the HLS tool and the optimization directives provided. In fact, creators of the aforementioned high-performance accelerators \cite{liu2016high, zhou2018rosetta, zhang2017high, li2019implementing, zhao-bnn-fpga2017, han2017ese, jason1} usually have years of experience with HLS. In many cases, due to the high complexity of the design, the interactions between the optimization options cannot be discovered by experienced HLS designers.

Such challenges inspire us to develop an automated design space exploration (DSE) tool to find the optimal configuration of the optimization directives. For these automated HLS DSE tools, there are several key challenges: The most important one is to reduce the number of invocations of the HLS synthesis process. The reason is that synthesizing a design usually takes several minutes, which means evaluating a large number of design points during the DSE process can be extremely time-consuming. While creating analytical models can be a viable approach to speed up the evaluation, the internal mechanisms of HLS are complicated and cannot be accurately modeled with simple analytical expressions. In many cases, to compute an accurate prediction, the analytical tools still need to perform complex dependency analysis and scheduling, which are also time-consuming. 

In addition, when designing an accelerator, the designers will need multiple options in terms of latency and resource trade-offs, so that they can find the best combination of design parameters that balance the performance and resource usage. For this reason, the DSE tool also needs to perform multi-objective optimizations, so that it not only finds the extreme design points with the lowest latency, resource, or power consumption but also attempts to find the Pareto efficient points in between.

Finally, due to the complex nature of the HLS process, the performance or resource usage of the synthesized hardware, as a function of the input design point, is highly nonlinear and multimodal, which means an optimization method can easily fall into local optima. Therefore, it is crucial for the DSE method to have the ability to escape local optima effectively.

In this paper, we present Chimera, a machine learning driven DSE tool for HLS that aims to solve the aforementioned challenges. This work has four major contributions in solving the aforementioned challenges:
\begin{enumerate}
    \item We enable multi-objective optimization with active learning, which uses the predictions from the machine learning (ML) model itself to create the training dataset. As a result, effective performance/resource models can be built with a significantly smaller number of evaluated samples, reducing the number of HLS invocations during the exploration process. It also enables information sharing between different exploration strategies.
    \item We leverage the evolutionary algorithm (EA) in conjunction with ML to achieve a higher probability of discovering Pareto optimal points in each step of the exploration. Comparing to proposing points randomly or from a simple probability distribution \cite{hypermapper,hypermapper2.0}, EA enables the DSE tool to incorporate more information from known design points.
    \item We develop a ``soft-boundary'' technique that selects points to explore in a probabilistic manner, making the optimization process less prone to poor initialization. Comparing to the hard ``pass-or-fail'' decision method, this technique accounts for the inaccuracies of the models and avoids being overly greedy.
    \item We utilize the Thompson sampling heuristic to create a hybrid method that adaptively switches between exploration and exploitation strategies, which greatly helps escape local optima. When trapped at local optima in the design space, the Thomson sampling heuristic will switch from the greedy exploitative techniques to exploratory technique to introduce new information to the dataset, so that the tool can escape from local optima.
\end{enumerate}

With these technical advancements, for the real-world benchmarks in the Rosetta benchmark suite \cite{zhou2018rosetta}, Chimera not only matches or surpasses the low latency design points found by human designers but also explore a whole Pareto curve that represents efficient combinations of latency and resources in less than 24 hours.

The following sections are organized as follows: Section II introduces the background of the techniques and algorithms used. Section III is dedicated to discussing the differences and advancements of Chimera compared to previous related works. Then, in Section IV, the overall flow of Chimera is described and discussed in detail. Section V describes the experiments conducted to evaluate the Chimera tool and present the results acquired with discussion. Finally, Section VI concludes this paper with the current limitations and future works.

\section{BACKGROUNDS}
\label{sec:backgrounds}
\subsection{Xilinx Vivado High-Level Synthesis}
The Xilinx Vivado HLS \cite{xilinx_vivado, vivado_intro} is a representative HLS tool that has been used in a wide range of research projects and real-world designs. It supports C/C++ to register-transfer-level (RTL) synthesis and provides a suite of optimization directives/pragmas. The combination of these directives/pragmas forms a large design space and the users usually need to explore the design space by evaluating many design points to find the optimal one. The pragmas/directives will be referred to as \emph{directives} from here on.

Among the types of directives, the \emph{loop directives} and \emph{array directives} are most commonly used. As the survey by Schafer and Wang \cite{hls_dse_survey} pointed out, the loop and array directives provide the most direct and fine-grained control of the synthesized hardware. Other similar works in this research area such as \cite{cilardo2015date}, \cite{linanalyzer}, and \cite{choi2018iccad} also focus on the same types of directives. In this work, we refer to these loop and array directives as \emph{tunable knobs}.

\begin{figure*}[h!]
    \centering
    \includegraphics[width=0.8\textwidth]{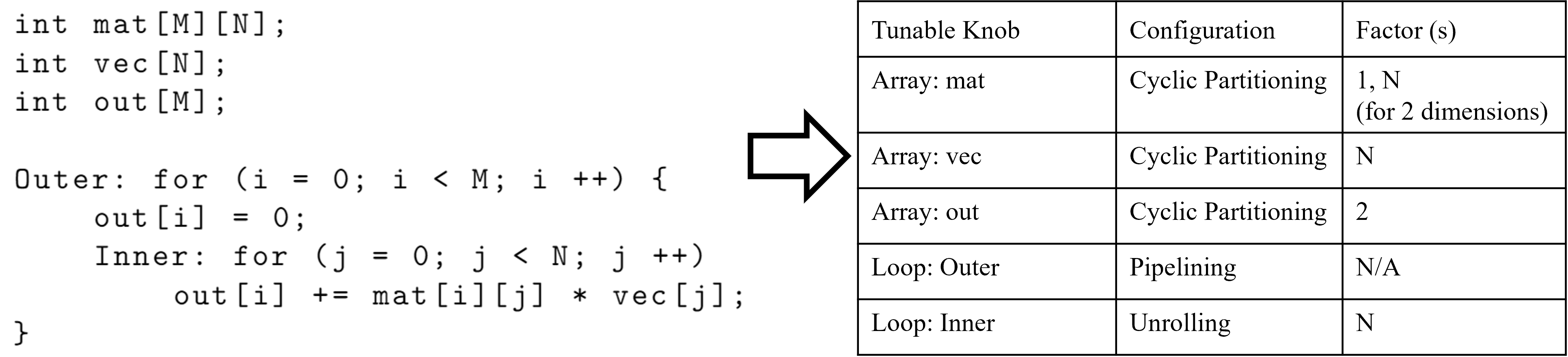}
    \caption{Example of a design point for a code snippet}
    \label{fig:design_point}
\end{figure*}

Each directive can have several configurations and a corresponding numerical factor. For instance, the loop directive has two configurations: \emph{pipelining} and \emph{unrolling}. The information of the directives used in this paper is summarized in Table \ref{tab:pragmas}. As an example, for the code snippet shown on the left of Fig. \ref{fig:design_point}, a design point for it can be summarized as the table on the right.

Further details about the directives are in the user manual for Vivado HLS \cite{vivado_user_guide}. 

\begin{table}[h!]
    \centering
    \vspace{-1mm}
    \caption{Summary of available options of the loop and array directives explored by Chimera}
    \begin{tabular}{|c|c|c|p{0.3\linewidth}|}
            \hline
            Tunable Knob & Type & Configuration & Factor Description \\
            \hline
            loop & pipelining & N/A & N/A\\
            \hline
            loop & unrolling & N/A & Specifies the number of iterations executed in parallel\\
            \hline
            array & partitioning & complete & N/A\\
            \hline
            array & partitioning & cyclic & Specifies the number of smaller arrays that the array is partitioned into\\
            \hline
            array & partitioning & block & Same as cyclic partitioning\\
            \hline
    \end{tabular}
    \label{tab:pragmas}
    \vspace{-2mm}
\end{table}

\subsection{Characteristics for the optimization problem in HLS DSE}
The design space exploration for an HLS design can be regarded as an optimization problem, and it has some unique characteristics that make it challenging to solve. 

One of the key characteristics is that, due to the high complexity of HLS synthesis, the synthesizable design points in the input design space might not be contiguous. In other words, there can be isolated 'islands', where the points in side the 'island' are synthesizable, but almost all points outside the 'island' that are close the boundary are not synthesizable. In such a case, the algorithms that explore by evaluating neighbors of known design points can hardly escape the island and fail to find the global optimal. Therefore, conventional optimization methods such as simulated annealing might not be effective for HLS DSE.

Another key challenge of HLS DSE is the complex interactions between the tunable knobs, that is, multiple pragmas usually has to be configured in specific combinations to improve the performance. For instance, the partitioning factor of an array might need to match the unrolling factor of the loop that accesses this array. Modifying any of them individually might lead to no improvement or even degradation in latency and resource consumption. Furthermore, depending on the size of the design, an HLS program can have a large number of tunable knobs, which means that finding the best combination of modifications by evaluating all the neighboring points exhaustively is not practical. Therefore, the optimization algorithm has to have the ability to extract hidden information from the known design points to explore the design space efficiently. Hence, we see the need for applying machine learning in this situation.

\subsection{Multi-Objective Optimization}
As the name suggests, in a multi-objective optimization problem, the algorithm needs to find the optimal solutions considering multiple conflicting objective functions. This type of optimization problem is common in hardware design processes. For an HLS design on FPGAs, there are three conflicting objectives that a designer has to optimize for: lower latency, lower resource usage, and lower power consumption. For example, a design with lower latency might use more hardware resources to parallelize the computation. On the contrary, a design that uses fewer hardware resources might have to serialize the computation, which leads to higher latency. Consequently, the designer must balance between latency, resource usage, and power consumption and find the optimal points that represent the best trade-offs between the objectives.

\emph{Solving an optimization problem for low latency and low resource usage is the main goal of Chimera}. Formally, the optimization problem we focus on in this paper can be formulated as the following: Denote the design space defined by the user for code $K$ with $\mathbb{R}_{K}$. Each design point $d_i$ within the design space $\mathbb{R}_{K}$ can be evaluated by the target black-box HLS tool to acquire a resource usage value $R(d_i)$ and a latency $L(d_i)$. Then we can define the objective as: Find the Pareto optimal set $\mathbb{P}_{K} \subseteq \mathbb{R}_{K}$, such that for any $p_j \in \mathbb{P}_{K}$, there is NO design point $d_i \in \mathbb{R}_{K}$ that satisfies $R(d_i) < R(p_j)$ and $L(d_i) < L(p_j)$ simultaneously.

\subsection{Decision Tree Learning}
The decision tree (DT) is a machine learning algorithm that uses a binary tree structure to represent the learned information. One of the main advantages of the decision tree-based models is their ability to learn nonlinear relationships with a relatively small dataset. This makes them suitable for modeling the performance and resource usage of HLS designs. As mentioned before, the performance of a design point cannot be predicted by linear models without complicated mathematical transformations. Moreover, the time-consuming synthesis process limits the amount of data that can be collected, resulting in a relatively smaller dataset.

However, the simple DT model usually suffers from overfitting issues; therefore, in this work, we choose to use Random Decision Forest (or Random Forest, RF) models to predict the performance and resource usage of the design points. The  RF algorithm was created based on DT to improve the overall accuracy \cite{breiman2001random}. The Random Forest (RF) is an aggregated learning algorithm that consists of a large number of weak decision trees. It applies the bootstrap aggregating (bagging) strategy, which makes the training process more resistant to overfitting compared to the simple decision tree model.

\subsection{Active Learning}
Active learning is a family of machine learning methods that mainly focuses on reducing the number of samples needed for constructing accurate models \cite{settles2009active}. It uses the model's own predictions to selectively label the samples that provide the most valuable information to the model, thus reduces the number of labeled training samples needed to construct an accurate model. It originated as a technique to improve the predictive power of a machine learning model with a fixed number of training samples \cite{cohn1996active}. It was later applied to the auto-tuning of program and compiler parameters in \cite{hypermapper, hypermapper_3dvision}. Recently, it was also leveraged by the Spatial language \cite{koeplinger2018spatial} as the backbone of its DSE engine and demonstrated great potential \cite{hypermapper2.0}.

In this work, we apply active learning by iteratively using the models to selectively add potentially Pareto efficient design points to the training samples and retrain the model with the updated dataset. In each iteration of this process, the newly added training samples help improve the accuracy of the models, which, in turn, is used to discover more Pareto efficient points effectively. Meanwhile, this training process can also be seen as a multi-objective optimization process, as we are exploring more Pareto efficient points as we construct the training dataset.

\subsection{Thompson Sampling}
Thompson sampling is a heuristic commonly used for multi-armed bandit problems \cite{russo2017tutorial}. For such problems, the algorithm faces multiple options that can potentially yield rewards and it needs to maximize the rewards it gains over multiple attempts. The naive strategy of simply choosing the option that has the highest expected reward in the history of attempts can be too greedy and can easily fall to local optima. This is because the better option might be underrated due to the lack of attempts. In other words, the uncertainty of the expectation is not taken into consideration, since a smaller number of attempts means the uncertainty in the expected reward of an option is higher.

In essence, the Thompson sampling learns the expected reward for multiple possible options, but it not only considers the option that yields the highest reward but also considers the uncertainties of the expectations, such that the exploration and exploitation are well balanced. This means that, even if some options have a lower expected reward in the history, it can still be selected for exploration purpose if there are much fewer attempts on it in the history.

For a DSE problem, a single exploration technique might not be effective in all scenarios. For example, the EA-based exploration can be overly greedy and prone to local optima, whereas the random exploration can be ineffective in discovering extreme points. Given these limitations, we use Thompson sampling to combine the strengths of multiple techniques, creating a hybrid technique that can be effective in all scenarios.

\vspace{-1mm}
\section{RELATED WORKS}
\vspace{-1mm}
There has been several previous attempts on automating DSE for HLS directives\cite{linanalyzer, autoaccel, choi2018iccad, COMBA}. However, these previous works are based on single-objective optimization methods, which means they only optimize for one objective, such as low resource or low latency. On the contrary, Chimera is based on a multi-objective optimization method that provides a wide range of design choices with different resource and latency trade-offs. This allows the designer to pick design points for different resource budgets easily when the design constraints change, without rerunning the whole DSE.

Furthermore, as \cite{hls_dse_survey} pointed out, previous works that use analytical models can become inaccurate as they depend on the expert knowledge about the constantly evolving HLS tools. Additionally, as noted in \cite{hls_dse_survey}, the HLS tools utilize heuristics that have limitations, which the analytical models cannot account for. Consequently, the estimation from an analytical model can differ significantly from the actual estimation from HLS in practice. In comparison, Chimera directly relies on the output from the underlying HLS tool. Thus, even if the generated hardware is not ideal due to the limitations of the HLS tool, the output of the exploration can still approach the best attainable performance in reality, instead of converging to a hypothetical optimal point given by the ideal analytical model. In addition, using an analytical model means migrating between different HLS tools requires major changes to the DSE tool. However, for Chimera, since the optimization method does not depend on the internal mechanisms of HLS, it can be easily migrated and extended to future HLS tools or HLS tools from other vendors. Finally, Chimera uses machine learning models to learn to avoid the un-synthesizable design points, which making it more practical to use in real-world scenarios. 

While some other previous works such as \cite{schafer2012, liu2013} also apply data-driven methods to enable multi-objective optimization, comparing to Chimera, \cite{schafer2012} only applied random sampling instead of active learning, which means it will require more samples to construct an effective model; \cite{liu2013} applied selective sampling for the training dataset but does not use the predictions from machine learning model and does not integrate evolutionary algorithm. Moreover, Chimera employs Thompson sampling to combine the strengths of various optimization methods and help escape local optima effectively, making it more efficient in exploring large design space. Finally, we demonstrates Chimera's capability on a realistic benchmark suit, whereas \cite{liu2013} evaluated the methodology on a single benchmark application with a limited number of tunable knobs.

\section{METHODOLOGY}
Chimera is an automated DSE tool dedicated to optimizing the directive configurations of C/C++-based HLS designs for FPGAs. As an automated DSE tool, in addition to the Vivado project source files, the user only needs to provide a comma-separated values (CSV) file containing the information of the tunable knobs. Currently, Chimera is targeting the Xilinx Vivado HLS environment, but it can also be generalized to other HLS toolsets as well.

\begin{figure}[h!]
    \centering
    \vspace{-3mm}
    \includegraphics[width=0.75\columnwidth]{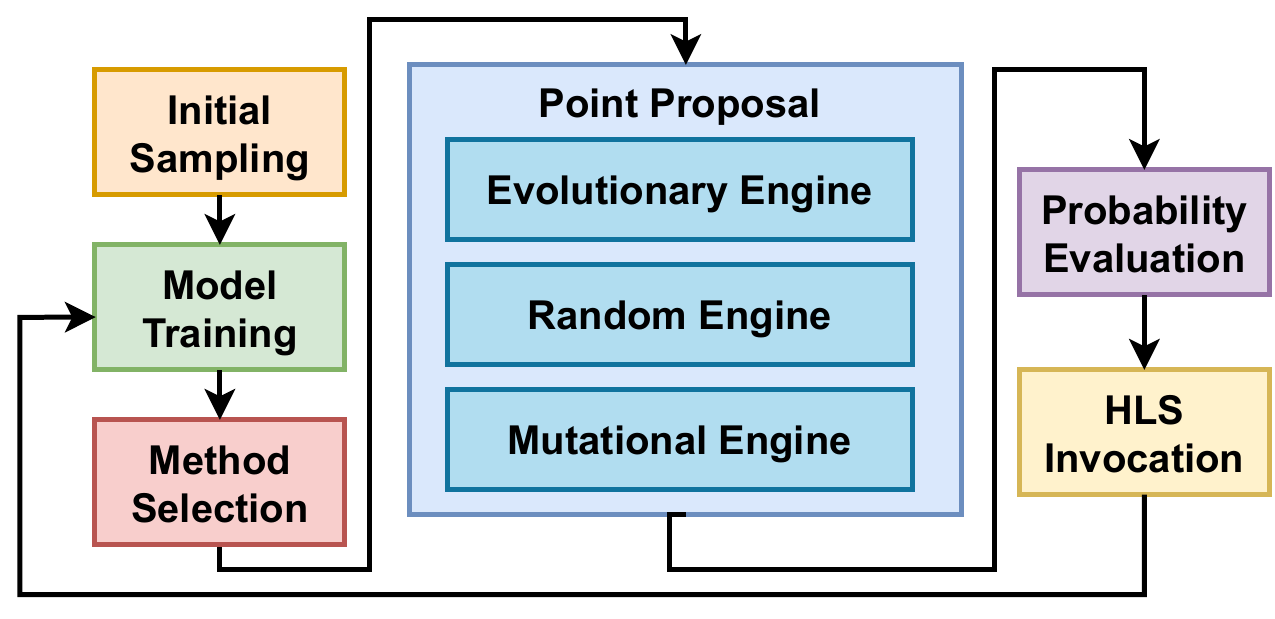}
    \caption{Overall flow and stages of the Chimera DSE process}
    \label{fig:overall_flow}
\end{figure}

Fig. \ref{fig:overall_flow} demonstrates the main stages and the overall flow of the DSE process. In the initial sampling stage, Chimera randomly evaluates a small set of design points in the design space. Then, in the model training stage, it trains the machine learning models with these initial samples. The models are used for predicting the resource usage and latency for all later stages of the DSE process. Once the models are trained, in the method selection stage, it uses the Thompson sampling heuristic to pick one of the three point proposal engines to propose the next design points to be potentially explored. After that, it enters the probability evaluation stage, where a probability will be assigned to the proposed design point based on its predicted performance and resource usage. This probability determines how likely this design point will be evaluated. Following that, if the new point is selected to be evaluated, the information of the new point will be passed to the HLS invocation stage, which will call the Vivado HLS synthesis tool. The results from HLS will then be added to the training dataset. It will also determine whether the evaluated design point is superior to the Pareto frontier and record it in the history of the point proposal engines. Finally, the tool will return to the model training stage of the loop, where the models are retrained. This loop will then repeat until the stopping condition is met.

\subsection{Initial Sampling}
In the initial sampling stage, Chimera will randomly select a set of design points from the design space, and evaluate the latencies and resource usages. In this stage, the machine learning models are not trained yet, hence will not be used to pre-select the design points for HLS. All evaluated design points will then be added to the dataset.

\subsection{Model Training}
For Chimera, the ML models are the keys to improve the efficiency of sampling the design space. Specifically, their main function is to predict the performance, resources, as well as the possibility of encountering errors for a design point. ML has been recently adopted by various design automation areas, and is especially effective for hardware co-designs on FPGAs \cite{AutoDNNchip, CoDesign, CoDesignSummary,EDD}. For this work, as mentioned in Section II, the Random Decision Forests were used for all the predictive models in Chimera, but other suitable ML models can also be used for a different DSE problem.

The input features to the models comprise directive configurations and factors converted to numerical data. For each of the tunable knobs, such as a loop or an array, there are two input features defined, namely, \texttt{config} and \texttt{factor}. The \texttt{config} corresponds to the configuration of a directive. For example, for the loops, the \texttt{config} feature can be ``pipelining'' or ``unrolling''. The \texttt{factor} is a numerical feature that corresponds to the unrolling or partitioning factor where applicable. The types of the directive, as categorical features, are converted to numerical variables with one-hot encoding before feeding into the models.

For the resource prediction, there are four separate models, which predict the hardware resource utilization in terms of Block RAM (BRAM), Look-Up Table (LUT), Flip-Flop (FF), and DSP, respectively. The output predictions are the proportions of the corresponding resource type consumed on the target FPGA. For the latency prediction, the input features are the same as the resource predictors, and the output is the predicted latency, which is the multiplication of clock period and the total number of clock cycles.

For a practical DSE tool, it is unrealistic to assume that all the design points can be synthesized without errors and within a reasonable amount of time. Therefore, in addition to predicting the resource and latency, Chimera also has a classification model dedicated to predicting the probability of encountering synthesis errors or timeout. The model learns from the design points that are not synthesizable or timed out during the exploration process, and make predictions for the newly proposed designs. We then use the predictions to reduce the risk of evaluating un-synthesizable design points. As a result, the time spent on waiting for timeout or error can be saved. In the actual implementation, the probability of timeout/error for the design point $P_{timeout}$ generated by the the classification model is used in the probability evaluation stage to determine the total probability of evaluating a point. Notice that, the quality of the explored design points is hardly affected by this, because the optimal points usually follow the design limitations of the HLS tool, so they are synthesizable and do not require excessive time to synthesize.

\subsection{Method Selection}
Chimera has three methods to propose new design points to explore starting from the initial sampling set: \emph{random}, \emph{evolutionary}, and \emph{mutational}. Among them, the evolutionary method is more effective in combining the beneficial directives from multiple known design points; the mutational method is more effective in discovering more extreme design points, such as points with lowest known latency or lowest known resource usages; the random proposal method employs random sampling is the most effective on introducing new information to the dataset. The three methods complement each other in terms of greediness and are tightly integrated by sharing the same set of machine learning models and sample population, making Chimera fundamentally different from a simple ensemble of exploration techniques.

Selecting the method for each iteration of exploration can be considered a beta-Bernoulli bandit problem, in which a positive reward $1$ will be assigned when a method finds a new Pareto non-dominated design point that pushes the Pareto frontier, and the algorithm needs to maximize the total reward in a finite number of attempts. Notice that we refer to the points as \emph{Pareto non-dominated} since they are the design points that are not surpassed by any other \emph{known} design point in every objective. However, these points are not necessarily the absolute Pareto optimal points, which can only be confirmed with exhaustive searches.

For our DSE problem, we refer to finding a new Pareto non-dominated point as a success, and otherwise a failure. In this type of problems, the distribution of the possible rewards for each of the options can be considered as a Bernoulli distribution; so, we apply the \emph{Thompson sampling} heuristic, in which the expectation of getting a reward for each of the options obeys a separate beta distribution where $\beta$ is the total number of attempts on the option and $\alpha$ is the number of successful attempts.

In each iteration of the exploration, a sample will be drawn from each of the three beta distributions, and the method corresponding to the sample with the highest value will be chosen. At the beginning of the exploration process, $\alpha$ and $\beta$ will both be initialized to $1$ for the distributions, giving the methods the same chance of being selected. As the exploration process goes on, the values of $\alpha$ and $\beta$ will increase according to the result of each attempt, and the distribution will be more concentrated around the mean reward $\alpha/(\alpha+\beta)$. Since the expected value from the beta distribution of an option is the same as the mean reward of it in history \cite{russo2017tutorial, forbes2011statistical}, in general, the method with a higher mean reward is more likely to be selected.

\begin{figure}
    \centering
    \includegraphics[width=0.6\columnwidth]{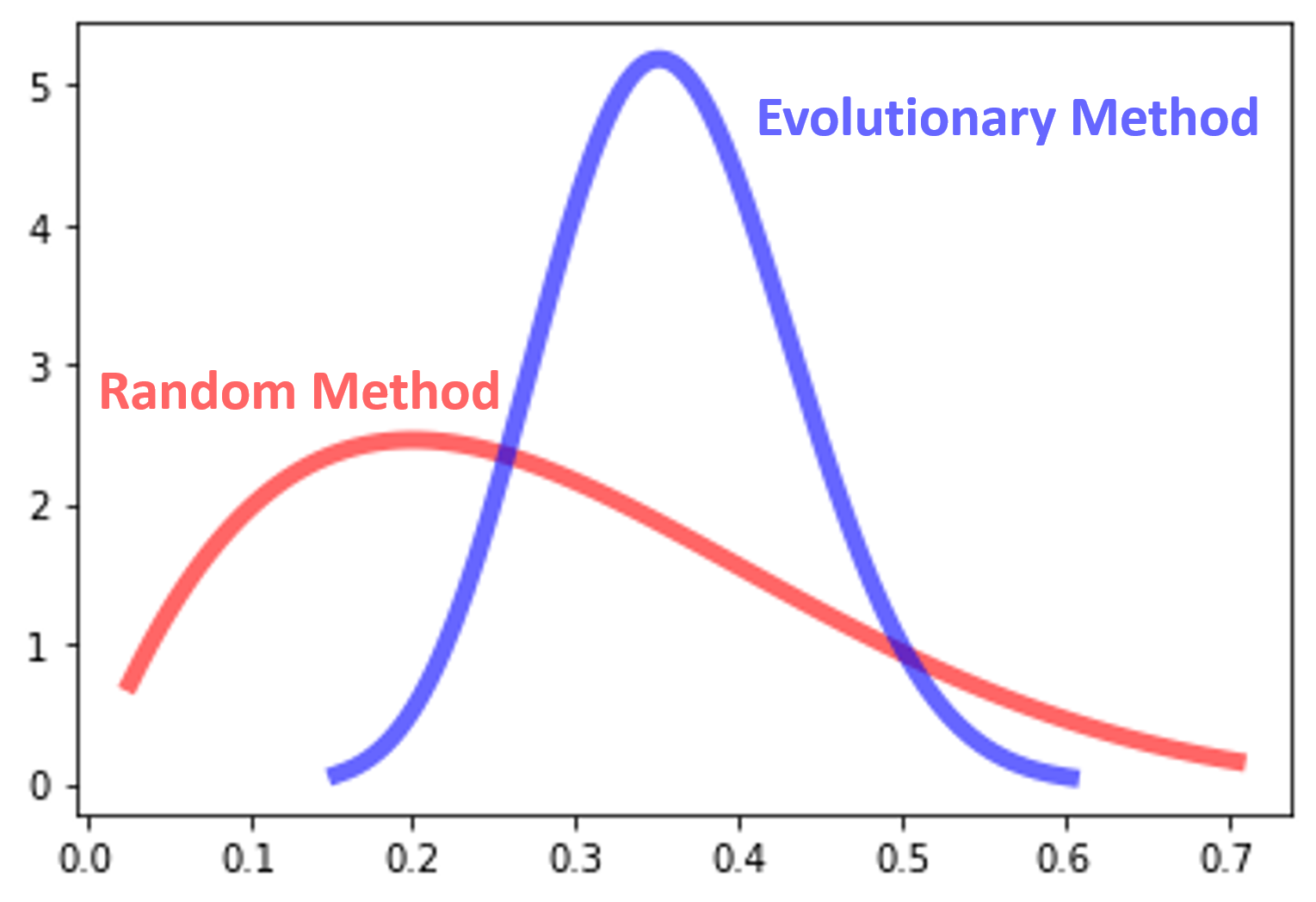}
    \caption{The PDF of the beta distributions for random and evolutionary method in the example}
    \label{fig:compare_beta}
\end{figure}

To understand how the Thompson sampling works in Chimera, consider a scenario where the random method is selected for five times with two successes among the past 30 attempts, and the evolutionary proposal method is selected for 25 times with 14 successes. In such a case, the probability density function (PDF) of the two methods is plotted in Fig. \ref{fig:compare_beta}. The red line represents the PDF of the beta distribution of the random method, and the blue line is for the evolutionary method. As the figure shows, although the mean reward from the history is higher for the evolutionary proposal, the PDF of the distribution for the random method is less concentrated. This means that the probability of getting a value higher than 0.5 is higher for the random method. For instance, when sampling the beta distributions, it is possible that the value for the evolutionary method is 0.45, whereas the value is 0.6 for the random method. So in such a case, the random method will be selected despite the lower mean reward in history. Such a situation can happen at the beginning of the exploration process. Without Thompson sampling, the evolutionary proposal method will always be selected, although it might have fallen into a local minima and the improvements to the Pareto frontier brought by the newly discovered Pareto non-dominated design points are minimal. In comparison, the Thompson sampling will select the random method occasionally, which means it allows less greedy exploration in the design space to help escape the local minima. In the experiments, we found that such a property is especially helpful in exploring the extreme design points with the lowest latency or lowest resource usage.

Notice that the effectiveness of a point proposal method is not completely fixed in the whole expanse of the exploration. In other words, the expected reward of each method can change as we collect more information from the design space. Therefore, Chimera only considers a limited part of the history of attempts. This allows it to actively change between exploration-oriented and exploitation-oriented engines during the exploration, which helps escape local optima.

\subsection{Point Proposal}
The point proposal stage generates the next design point to be explored. Therefore, the effectiveness of this stage ensures that the point explored can provide the most valuable information to the models and most likely to be Pareto non-dominant.

\subsubsection{Random Proposal Engine}
The random proposal engine is the simplest and the least greedy point proposal engine since it generates a random directive for each of the tunable knobs. But while being mostly random, the basic design rules of the HLS tool are still followed. For example, to avoid synthesis error, we apply the same type of partitioning on the dimensions of an array.

\subsubsection{Evolutionary Proposal Engine}
The evolutionary proposal engine proposes a new design points using the evolutionary algorithm. It creates a population of candidate designs that will evolve according to a fitness function, which defines the objective of the optimization process. Each individual in the population will have a certain genotype, which determines the phenotype of the individual. For our DSE problem, the genotype comprises the design parameters of the design point, and the phenotype is the performance and resource usage of the design point. The outline of the evolutionary proposal engine is shown in Algorithm \ref{alg:evo}.

\begin{algorithm}[t]
\KwData{Dataset of evaluated design points $\mathbb{D}$}
\KwIn{Number of families $n_{families}$ and number of offspring per family $n_{offspring}$}
\KwResult{Proposed design point $d_{prop}$ and the probability of evaluation $p_{eval}$}
\SetKwFunction{GetPopulation}{GetPopulation}
\SetKwFunction{Mutate}{Mutate}
\SetKwFunction{Crossover}{Crossover}
\SetKwFunction{GetProbEval}{GetProbEval}
\SetKwFunction{RandomSelection}{RandomSelection}

\emph{Initialize set of candidate design points $\mathbb{D}_{candidate}$}\;
\emph{Initialize set of probabilities of evaluation for the design points} $\mathbb{P}_{candidate}$\;
\tcp*[h]{\emph{Set the population}}\\
$\mathbb{D}_{pop}\leftarrow$GetPopulation($\mathbb{D}$)\;
\tcp*[h]{\emph{Generate offspring}}\\
\For{$i\leftarrow0$ to $n_{families}$}{
    $d_{father}^{(i)}\leftarrow$ \RandomSelection{$\mathbb{D}_{pop}$}\;
    $d_{mother}^{(i)}\leftarrow$ \RandomSelection{neighboring points on Pareto frontier}\;
    \For{$j\leftarrow0$ to $n_{offspring}$}{
        $d_{offspring}^{(i,j)}\leftarrow$ \Crossover{$d_{father}^{(i)}$, $d_{mother}^{(i)}$}\;
        $p_{offspring}^{(i,j)}\leftarrow$ \GetProbEval{$d_{offspring}^{(i,j)}$}\;
        $d_{mutant}^{(i,j)}\leftarrow$ \Mutate{$d_{offspring}^{(i,j)}$}\;
        $p_{mutant}^{(i,j)}\leftarrow$ \GetProbEval{$d_{mutant}^{(i,j)}$}\;
        $\mathbb{D}_{candidate}\leftarrow\mathbb{D}_{candidate}\cup\{d_{offspring}^{(i,j)}, d_{mutant}^{(i,j)}\}$\;
        $\mathbb{P}_{candidate}\leftarrow\mathbb{P}_{candidate}\cup\{p_{offspring}^{(i,j)}, p_{mutant}^{(i,j)}\}$\;
    }
}

\tcp*[h]{\emph{Select the design with highest probability of evaluation}}

$p_{prop}\leftarrow\max{(\mathbb{P}_{candidate})}$\;
$d_{prop}\leftarrow$ $d\in\mathbb{D}_{candidate}$ corresponds to $p_{prop}$.
\caption{Outline of the Evolutionary Proposal Engine}
\label{alg:evo}
\end{algorithm}

\begin{figure}[h!]
    \centering
    \vspace{-10mm}
    \includegraphics[width=0.8\columnwidth]{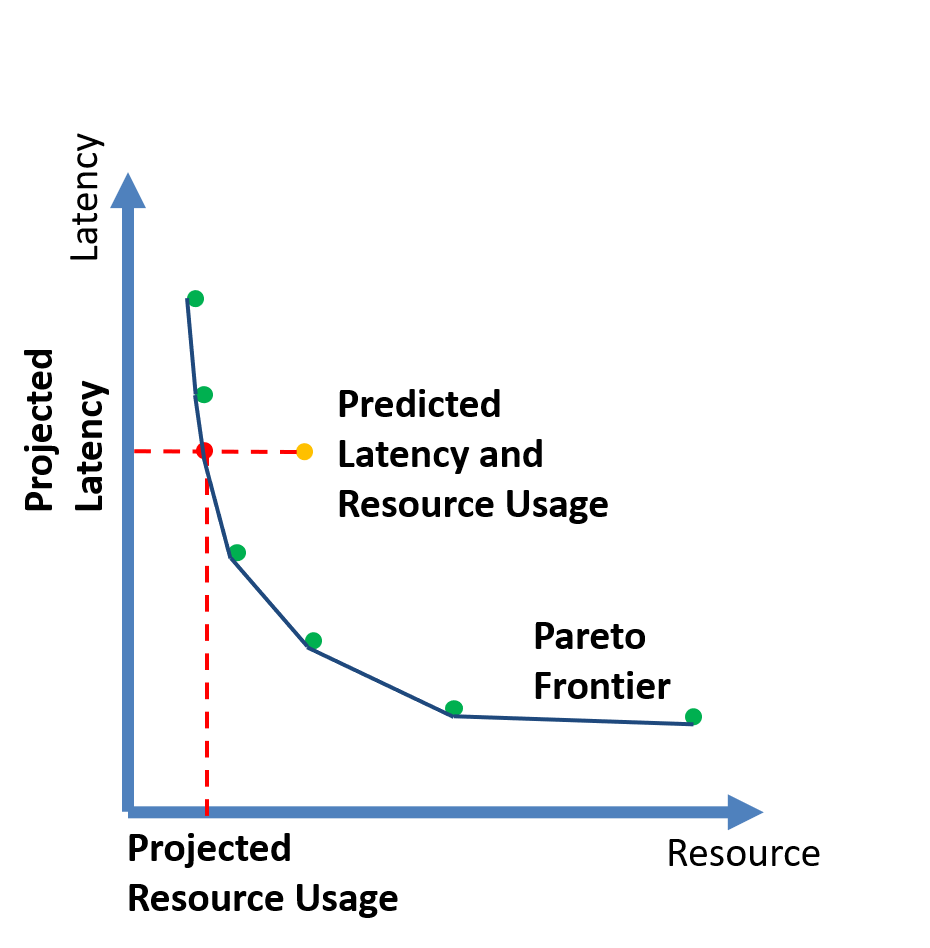}
    \vspace{-3mm}
    \caption{Example of the latency-wise projection of a design point on the Pareto frontier}
    \label{fig:latency_wise_projection}
\end{figure}

As Algorithm \ref{alg:evo} shows, the engine will first form a population with known design points that are close to the Pareto frontier, which is done by comparing the resource usage of a design point with the projected point on the frontier. As shown in Fig. \ref{fig:latency_wise_projection}, the projection is latency-wise and the resource of the projected point is found by linear interpolation. Within the population, it then randomly selects several design points as the ``fathers''. Each of the selected points will then breed with one of the neighboring design points on the Pareto frontier. The crossover and mutation happen during the breeding process, which generates several un-mutated offspring and mutants. Then, in the \texttt{GetProbEval} function, the surrogate ML models will be used to predict the latency and resource usage of these candidate design point, and a quality score will be calculated for each of them based on its distance to the Pareto frontier and whether it is feasible. This score is the same as the probability of evaluation to be described in Section 3.5. Afterward, they will be added to a list of candidate points and their corresponding quality scores (the probability of evaluation) will also be recorded. Finally, the candidate design points will be sorted by their quality score and the one with the highest score will be selected to pass to the next stage. The quality score, that is, the probability of evaluation, will also be passed to the next stage.

Comparing to the ordinary EA, the main strength of our variant is that active learning is incorporated to preselect the offspring before evaluating them, which reduces the total number of HLS invocations needed to find new Pareto non-dominated points. Specifically, the ML models are used to determine the probability of evaluation in the \texttt{GetProbEval} function, which, in turn, determines whether a design point will be proposed for evaluation. Also, the strategy of breeding is improved comparing to random breeding, such that the evolution is more effective for multi-objective DSE.

\subsubsection{Mutational Proposal Engine}
The mutation proposal method first randomly selects a point on the Pareto frontier constructed so far and mutates it several times to generate a set of mutants. Then, the mutants with the highest predicted probability of evaluation will be selected. 

It can be seen as a greedier version of the evolutionary method since it only applies mutation to the design points on the Pareto frontier. Its greediness means that it can be more effective in discovering extreme design points, but also more likely to fall into local optima.

\subsubsection{The Exploration-Exploitation Loop}
The three engines do not work as three individual search methods; instead, they work closely in conjunction with each other. While they have different internal algorithms, they share the same set of machine learning models and population of design points, making Chimera fundamentally different from a simple ensemble of exploration techniques.

\begin{figure}[h]
    \vspace{-3mm}
    \centering
    \includegraphics[width=0.9\columnwidth]{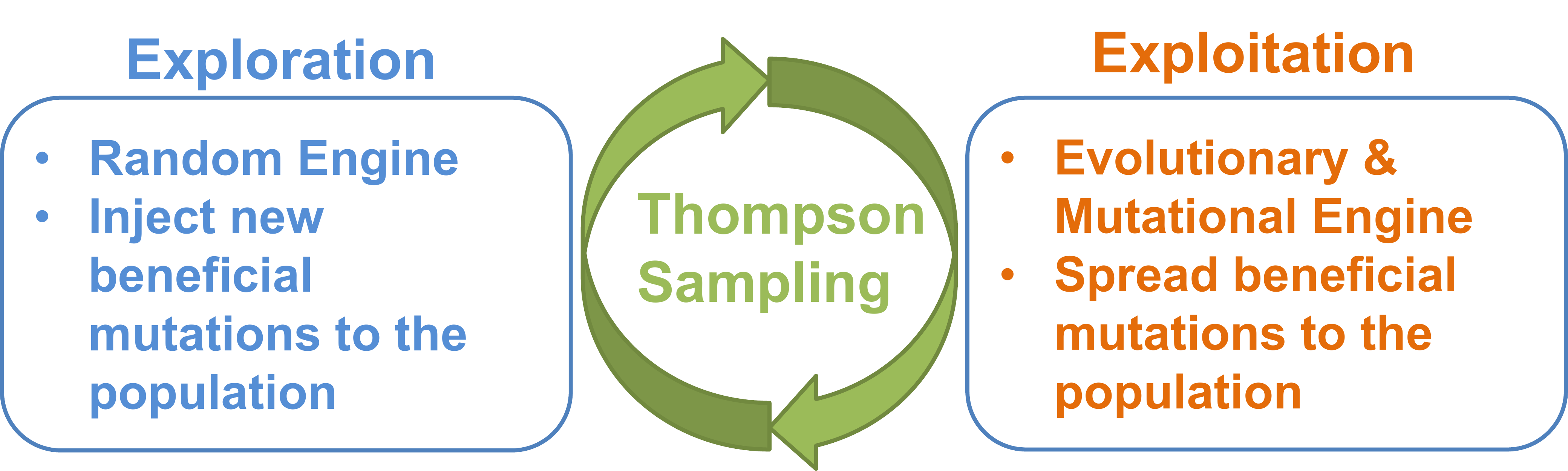}
    \vspace{-1mm}
    \caption{Summary of the ``exploration and exploitation'' loop}
    \label{fig:exp_to_exp}
    \vspace{-2mm}
\end{figure}

As mentioned before, for each exploration iteration, the point proposal method used is determined by the method selection stage. As illustrated in Fig. \ref{fig:exp_to_exp}, by combining the three methods with Thompson sampling, the exploration process can be seen as a conceptual loop of ``Exploration-Exploitation (E-E)'':

\emph{Exploration}: After the first few steps of the DSE, the optimization process can potentially fall into a local optimum where crossover and mutations introduced by the evolutionary and mutational engines fails to find new Pareto non-dominated points. During this period, the random engine will be more likely to be selected by the Thompson sampling. Since it can introduce multiple mutations at once, with the help of the machine learning models, the random engine can simultaneously generate several mutations that have a beneficial combined effect, which is crucial to escaping local optima. However, notice that the design points found by the random engine do not necessarily be Pareto non-dominated; the random engine is effective as long as it can introduce beneficial mutations to the population. Also, this does not mean that using the random engine alone will be as effective, because its effectiveness is dependent on the quality of the machine learning models, whose high accuracy relies on the points explored by the other two methods.
    
\emph{Exploitation}: Once the new beneficial mutations are introduced to the population, they will be rapidly spread to other individuals in the population by the evolutionary engine, and the mutational engine will find more extreme points with the updated population. This process, in turn, generates more valuable information about the interactions between the mutations, which can help improve the ML models. After the exploitation period, when almost no new Pareto non-dominated points can be proposed with the evolutionary and mutational engines, the tool will start a new exploration period, and the next iteration of ``exploration-then-exploitation'' will begin. Unlike the epsilon greedy method that only has one “exploration-then-exploitation” loop with a fixed boundary, our method can infinitely continue to further improve the result of exploration.

\vspace{-4mm}
\subsection{Probability Evaluation}
One of the main differences between a DSE engine for FPGA and a conventional compiler auto-tuning tool is that the DSE engine has to consider the resource constraints of the FPGA platforms. Previous work \cite{hypermapper2.0} proposed an ad hoc method that adds a separate classification model to predict whether a design point will satisfy the resource constraints with a fixed threshold. However, this method could be overly greedy since it assumes that the model is fully accurate. For example, since many of the high-performance designs will use close to 100\% of the BRAM and DSP resource on the target FPGA, if the feasibility prediction is slightly inaccurate, many of the potential high-performance design points will be predicted to be infeasible and discarded. In addition, when determining whether a design point is worth being evaluated, previous works directly reject points that are predicted to be worse than the Pareto frontier, setting an abrupt decision boundary in the design space, which prohibits ``uphill'' exploratory movement. As \cite{autoaccel} pointed out, such a method assumes that the initial samples are directly in the neighborhoods of the global optima, so that the model can directly capture the structure of the whole design space. As a result, it is prone to poor initialization. Considering these limitations, Chimera accounts for the imperfection of the predictive models and use a probability-based ``soft-boundary'' approach that does not directly reject the potentially valuable design points, which allows for more exploratory ``uphill'' moves that help avoid falling into local optima and makes it more immune to poor initialization.

Instead of using a simple classification model to predict the feasibility, we use the predictions from the ML models and the Pareto frontier to determine the probability that the given design point is to be evaluated. This probability is defined based on the comparison with the projected point on the Pareto frontier and the resource usage, combined with several user-defined hyper-parameters. The design points that consume higher resources compared to the points with the same latency on the Pareto frontier, or are predicted to use higher than 100\% of resource will have a lower probability of being evaluated. The details can be summarized with the following equations:

\begin{align}
    \nonumber  P_{budget} = &1 - \min(1, \max(0, (r_{LUT}-1)) \\
                + &\max(0, (r_{FF}-1)) + \max(0, (r_{DSP}-1))\\
                 \nonumber + &\max(0, (r_{BRAM}-1))) \\
    R_{pred} = &W_{LUT} \cdot r_{LUT}+W_{FF}\cdot r_{FF}+ \\
                 \nonumber &W_{DSP} \cdot r_{DSP}+W_{BRAM}\cdot r_{BRAM}\\
    P_{pareto} = &1 - \min(1, \max(0, \frac{R_{pred} - \delta \cdot R_{pareto}}{R_{pareto}}))\\
    P_{eval} = &P_{budget} \cdot  P_{pareto} \cdot  (1-P_{timeout})
\end{align}

In the equations above, $r_{[resource]}$ are the proportions of consumed resources for the corresponding type. They are calculated by dividing the \emph{predicted} amount of resource consumption by the amount available. For a feasible design, they range from $0$ to $1$ for all four resource types. Equation (1) demonstrates the probability $P_{budget}$ associated with the resource constraints (budgets). It is calculated by subtracting $1$ with the sum of proportions of overuse for four types of resources. Thus, for feasible design points, the $P_{budget}$ is 1. For a potentially infeasible design, the $r_{[resource]}$ will be greater than 1, so the $P_{budget}$ will be penalized and become smaller than 1. 

As shown in equation (2), $R_{pred}$ is the weighted sum of the proportions of consumed resources, and $W_{[resource]}$ is the user-defined weight for the corresponding resource type. $P_{pareto}$ is the probability associated with the Pareto optimality, as shown in equation (3). In this equation, $R_{pred}$ is the weighted sum calculated in equation (2) and $R_{pareto}$ is the resource usage of the projected point on the Pareto frontier corresponding to the proposed design point, similar to the projection described in Section 3.4.2 and illustrated in Fig. \ref{fig:latency_wise_projection}. $\delta$ is a user-defined hyper-parameter that can be adjusted to compensate for the inaccuracies in resource usage predictions. Notice that we can set different $\delta$ values for design points proposed by different point proposal engines. 

For a proposed design point, the more resources it uses, the lower the probability it has. If a proposed design point uses less resources comparing to the projected point, its $P_{pareto}$ will be 1. Finally, $P_{eval}$ is the total probability of evaluating the proposed design point. In addition to $P_{budget}$ and $P_{pareto}$, the probability of timeout from the timeout prediction model is also considered.

\subsection{HLS Invocation}
The HLS invocation stage is for executing the HLS synthesis tool and collecting the results from the reports generated. When a design point enters the HLS invocation stage, the tool will first generate a Tcl script that sets the directives accordingly, then launch the HLS synthesis using the script and extract the results from the synthesis report when the HLS tool exits. In the meantime, it monitors the total runtime of the HLS synthesis and report an error if the time limit is exceeded or the HLS tool exited abnormally. Once the results are acquired, it will update the datasets.

It is worth mentioning that, for designs that has loops with variable boundaries, the Vivado HLS tool will not be able to estimate the latency directly from synthesis. In such cases, there are two possible solutions: One is to use the \texttt{loop\_tripcount} directive to manually provide representative values for the loops with variable trip counts, such that the HLS tool can use the information to estimate the latency; the other method is to simulate the execution of the hardware with C-RTL co-simulation. The first method is suitable for cases where the trip count can be estimated by analyzing the code structure or can use C-simulation to acquire the trip count information. The second method is more suitable for cases where precise estimation is needed and the co-simulation can be done in a relatively short time.

\section{EXPERIMENTS AND RESULTS}
We evaluate the efficacy of the Chimera DSE tool with the Rosetta benchmark suite \cite{zhou2018rosetta}. Specifically, it is tested on the six benchmarks in Rosetta that are dedicated to the Vivado HLS environment.

\begin{figure*}[p!]
    \centering
     \begin{tabular}{cc}
        \begin{subfigure}[b]{0.9\columnwidth}
         \centering
         \includegraphics[width=\columnwidth]{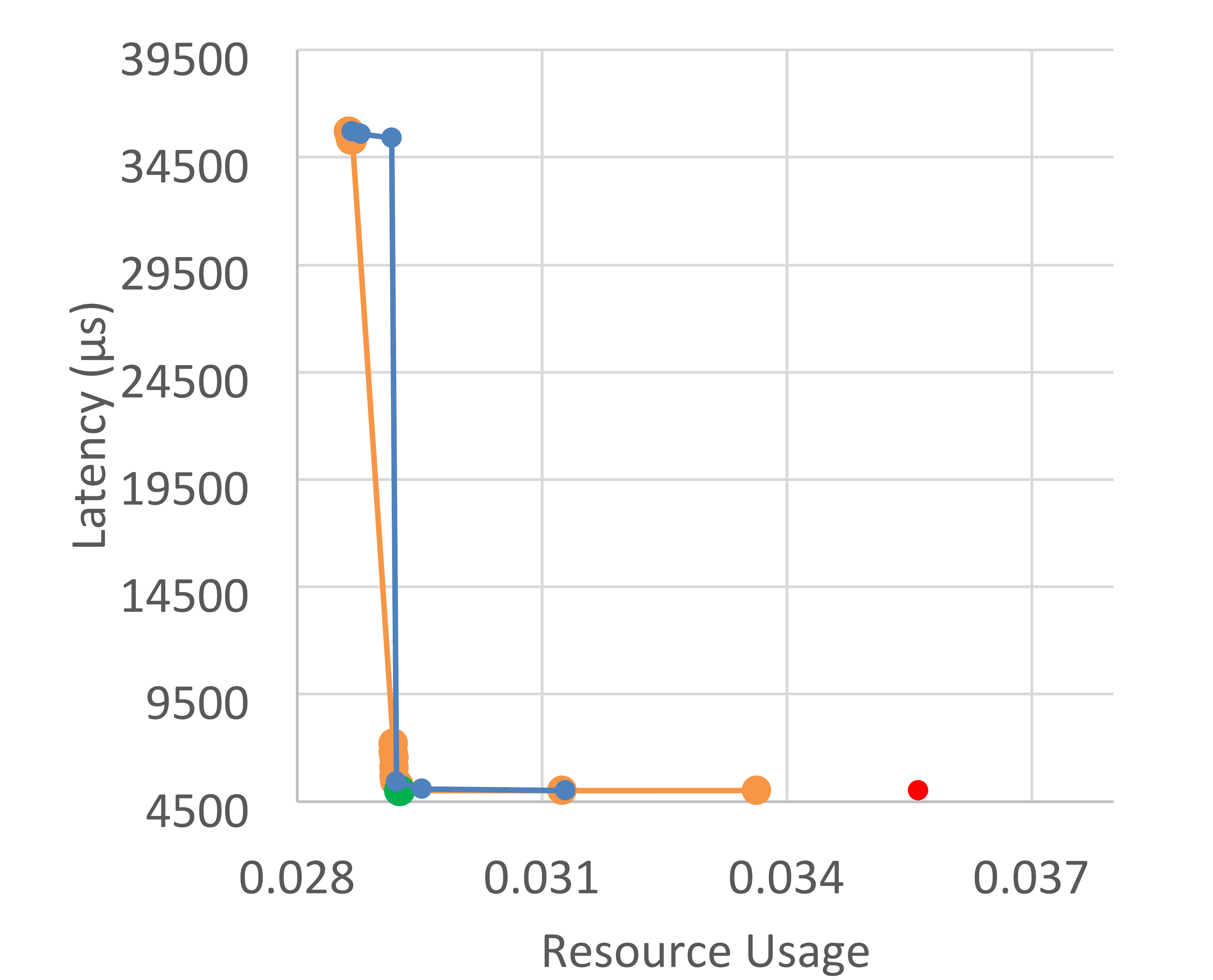}
         \caption{3D Rendering}
         \label{fig:results_orig_3dr}
        \end{subfigure} & 
         \begin{subfigure}[b]{\columnwidth}
         \centering
         \includegraphics[width=0.9\columnwidth]{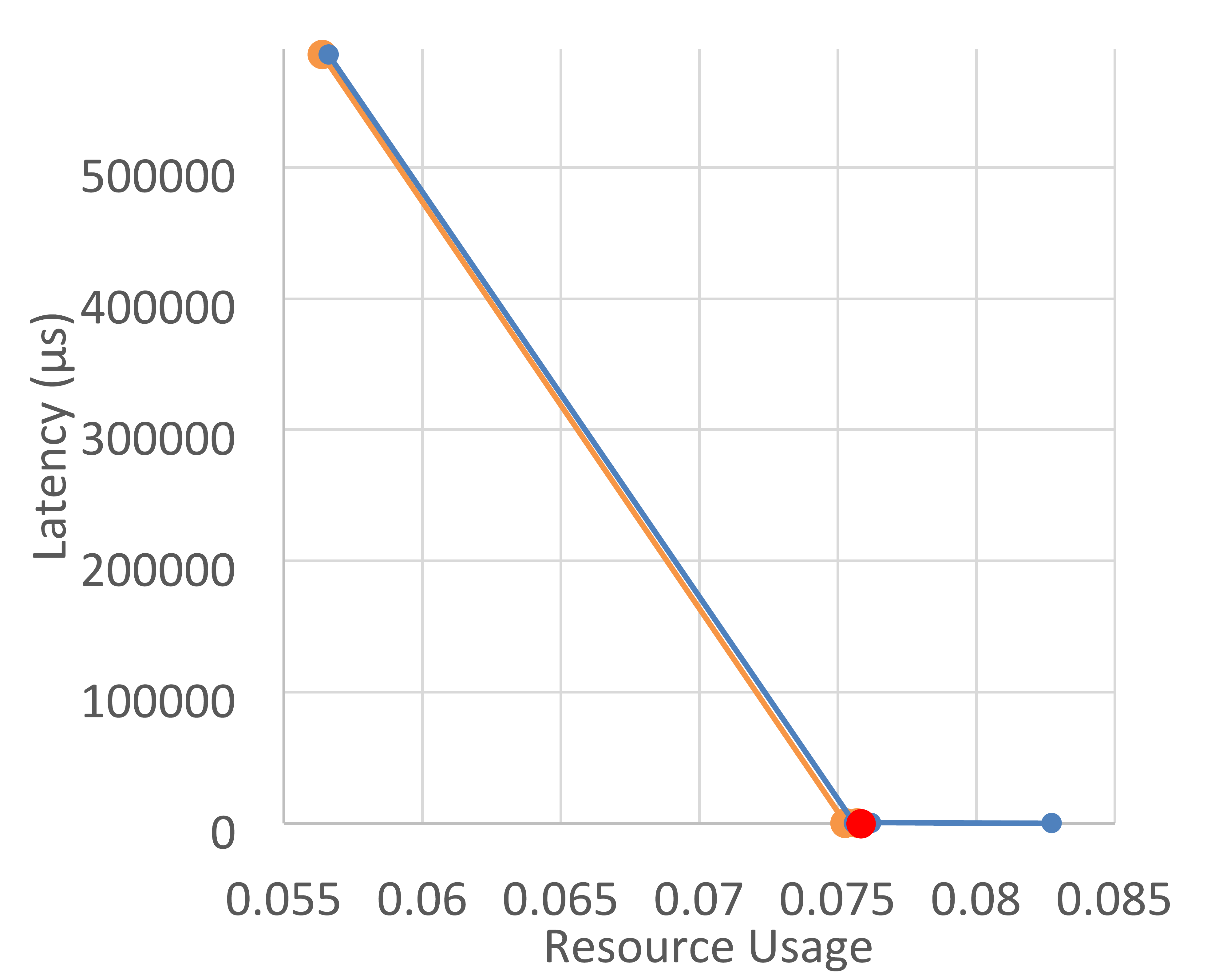}
         \caption{Binary Neural Network}
         \label{fig:three sin x}
        \end{subfigure} \\
        \begin{subfigure}[b]{\columnwidth}
         \centering
         \includegraphics[width=0.9\columnwidth]{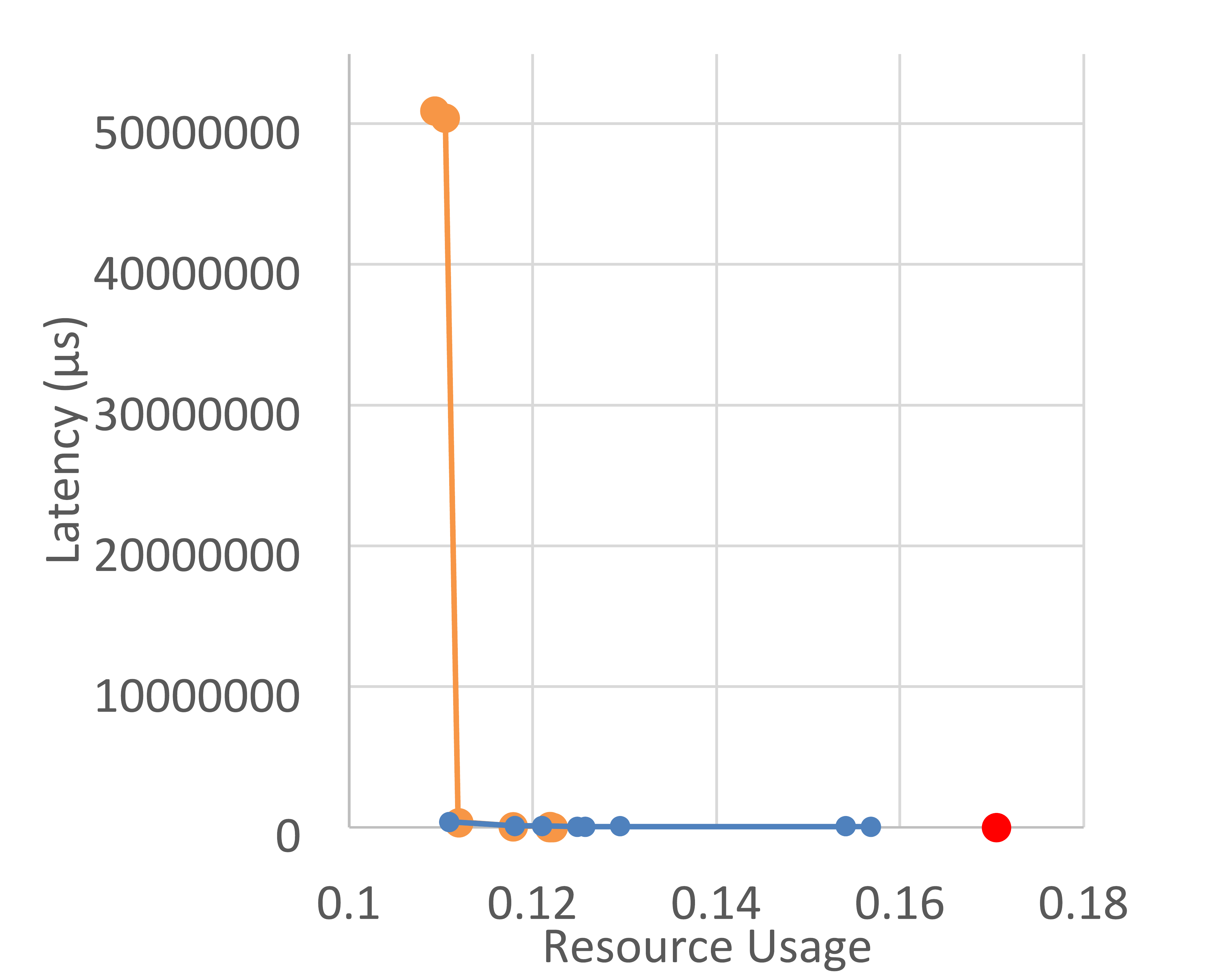}
         \caption{Digit Recognition}
         \label{fig:three sin x}
        \end{subfigure} &
        \begin{subfigure}[b]{\columnwidth}
         \centering
         \includegraphics[width=0.9\columnwidth]{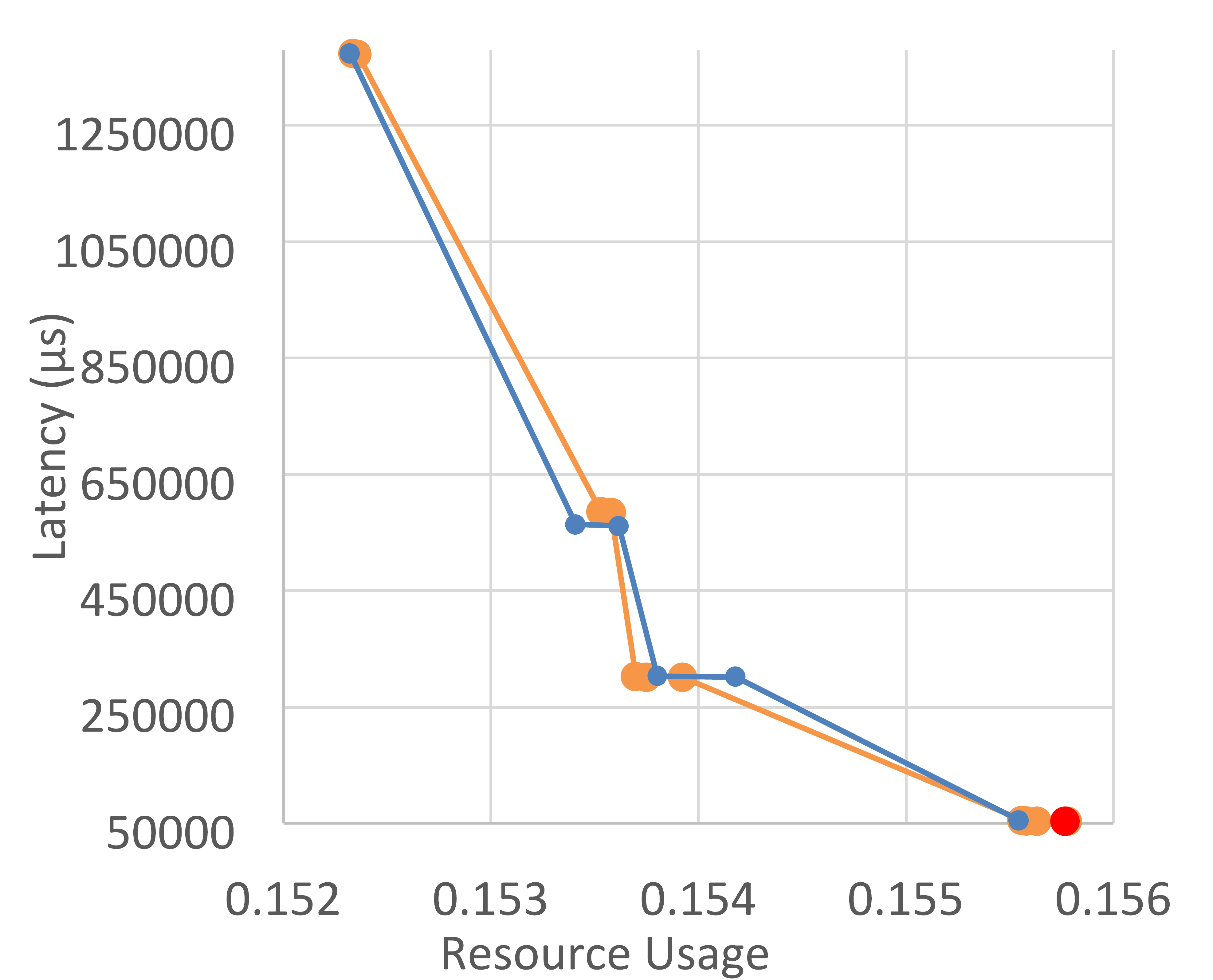}
         \caption{Face Detection}
         \label{fig:y equals x}
        \end{subfigure} \\
        \begin{subfigure}[b]{\columnwidth}
         \centering
         \includegraphics[width=0.9\columnwidth]{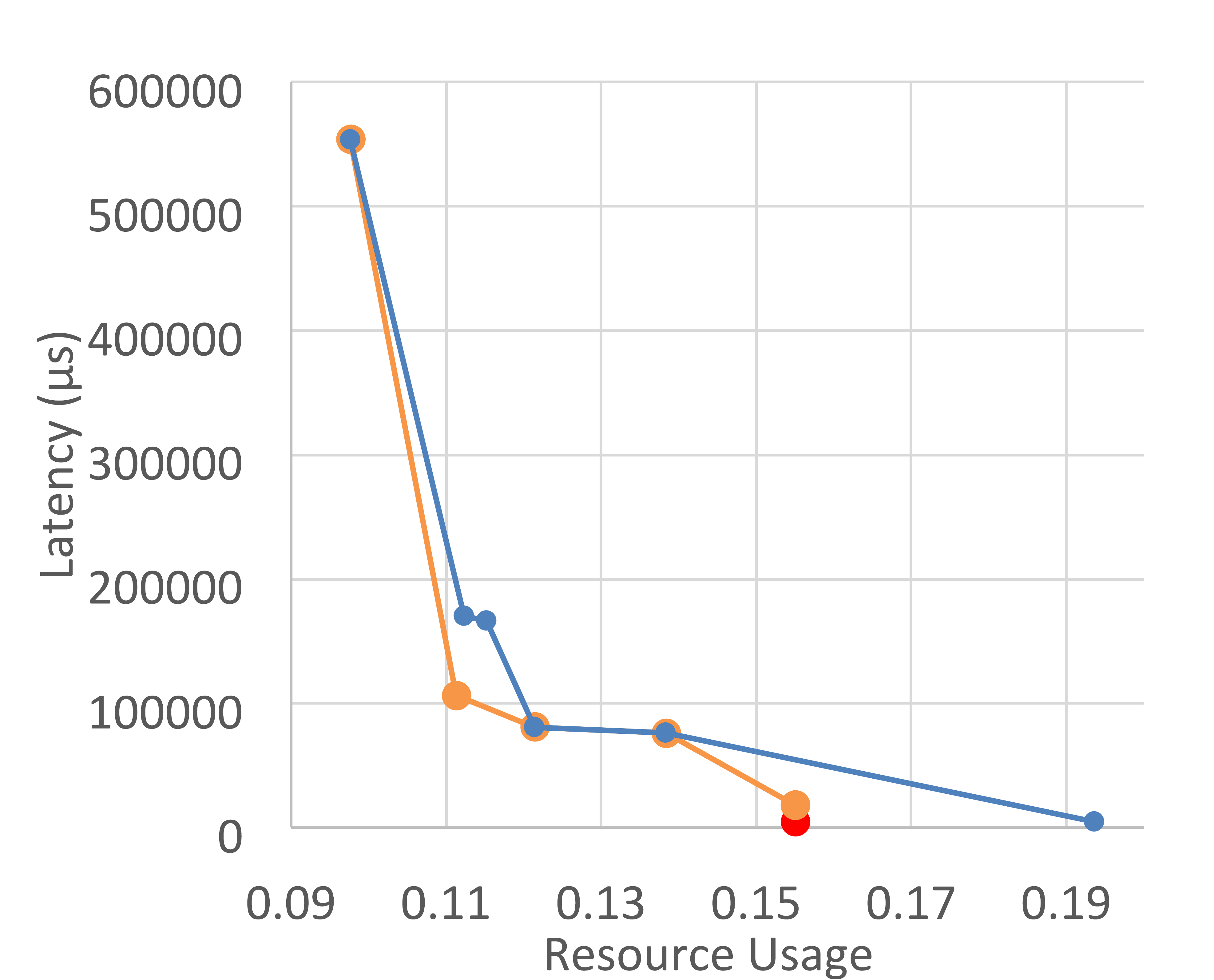}
         \caption{Optical Flow}
         \label{fig:three sin x}
        \end{subfigure} &
        \begin{subfigure}[b]{\columnwidth}
         \centering
         \includegraphics[width=0.9\columnwidth]{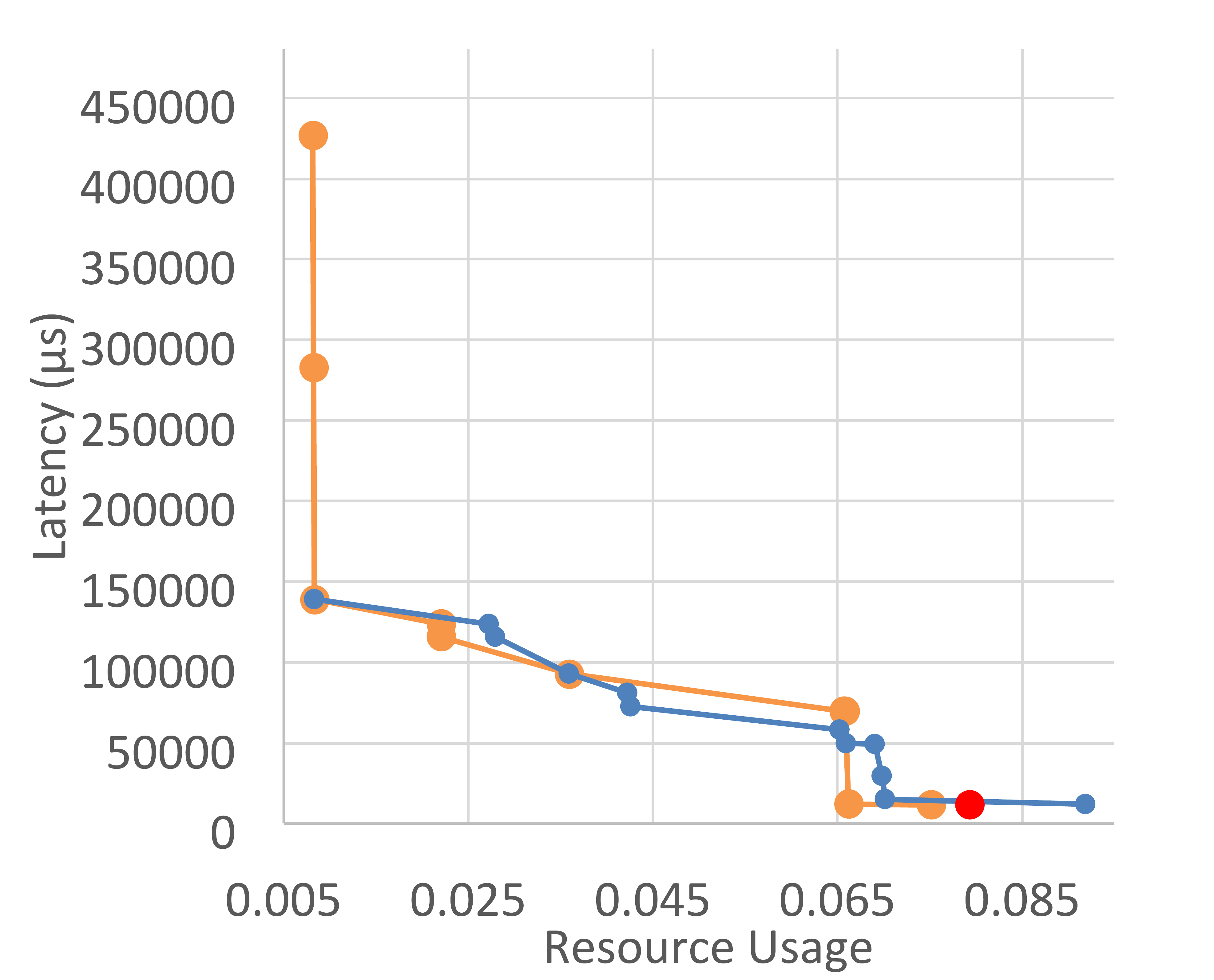}
         \caption{Spam Filtering}
         \label{fig:three sin x}
        \end{subfigure}
     \end{tabular}
     \caption{Pareto frontier and the hand-tuned design point}
     \label{fig:results}
     \clearpage
\end{figure*}

\begin{table}[h]
\vspace{-1mm}
\caption{Summary of the benchmark applications in the Rosetta benchmark suite}
\vspace{-1mm}
\centering
\begin{tabular}{|c|p{0.6\linewidth}|}
\hline
Name               & Description \\
\hline
3D Rendering (3DF)      & Renders 2D triangles into a 3D mesh model \\
\hline
BNN                & Binarized deep neural network \\
\hline
Digit Recognition (DR)  & Recognize digits using $k$-NN algorithm \\
\hline
Face Detection (FD)    & Detect faces using Haar cascade classifiers\\
\hline
Optical Flow (OF)      & Compute optical flow for a set of images \\
\hline
Spam Filtering (SF)    & Use a perceptron model to filter spam emails\\
\hline
\end{tabular}
\label{tab:rosetta}
\end{table}

The information of the benchmarks is summarized in Table \ref{tab:rosetta}. The HLS tool used for testing is Xilinx Vivado HLS 2019.2 and the experiments are conducted on a desktop workstation with an AMD 3900X CPU running at 4.1 GHz and 16GB DDR4-2133 memory.

There are several benefits of using the Rosetta benchmark suite to test a DSE tool for HLS. A major one is that the Rosetta benchmark suite consists of a set of real-world applications written for HLS, which have a higher level of complexity and larger numbers of tunable knobs. Using real-world HLS designs also means that there are more complex interactions between the tunable knobs. Some of these interactions can be hard to discover for human designers. Another key reason is that the time needed to synthesize a benchmark in it is longer, meaning that the importance of the sampling efficiency of the DSE tool is more pronounced. On the contrary, benchmark suites such as PolyBench \cite{pouchet2012polybench}, used by \cite{choi2018iccad, linanalyzer}, consist of a set of micro-kernels, which are relatively simple and cannot fully demonstrate the capability of a DSE tool. Additionally, Rosetta's benchmark applications are already hand-tuned by experienced designers, which provide a baseline for comparison.

In the experiments, we added loops and arrays that are nontrivial to tune to the set of tunable knobs. They not only include all tunable knobs at which the human designer inserted pragmas but also expands to other arrays and loops that have potential impacts on resource usage and latency. This enables Chimera to discover relationships between the directives that the human designers did not find, in order to find design points that are superior to the hand-tuned design. Notice that it is possible to include all loops and arrays of an HLS design in the CSV file and use Chimera to explore all possible design points. But in reality, similar to \cite{choi2018iccad}, we follow the basic design rules of HLS described in the Vivado user guide \cite{vivado_user_guide} to prune the design space by selectively adding arrays and loops to the set of tunable knobs. For example, loop layers with a large inner loop that also contains complicated operations are excluded, because unrolling or pipelining the outer loop will cause the inner loop to be completely unrolled, which leads to difficulties in scheduling. This kind of pruning can be done by a human designer with basic knowledge of HLS. Nevertheless, if a longer exploration time is allowed, the user does not need to prune the search space, since Chimera can learn to avoid the design points that take an extremely long time to synthesize.

Since the key target of Chimera is to be practical on real-world HLS designs, for the benchmarks in Rosetta, we aim to finish the exploration overnight on a typical workstation. Therefore, based on the preliminary profiling results on the benchmarks, we limit the total number of points to 170 for each run, such that the DSE for a benchmark can be finished within 24 hours. 

Despite the various design space sizes and complexities of different benchmarks, we found that 170 design points is sufficient for the benchmarks to converge to a stable Pareto frontier. Apparently, some benchmarks has relatively lower complexity and converges earlier, but for simplicity we set the explorations to end on the same number of design points. Such a result demonstrates the practicality of Chimera in real-world scenarios. For example, this enables the designers to run daily regressions with new design changes overnight, which means no engineer time is wasted. Among the 170 points, 20 points are the random samples evaluated in the initial sampling stage and 150 points are explored with the DSE algorithm. Notice that these numbers can be further tuned for better results for specific applications if needed. Also, due to the random nature of the algorithm, the result can vary slightly from run to run.

To acquire the Pareto curve, the resource usage value is calculated as a weighted sum of the proportion of consumed resources for the four types of resources. In the experiments, the weights are 0.4 for the BRAM and DSP, and 0.1 for LUT and FF. This is because the DSP and BRAM resources are much more scarce than LUT and FF on a typical FPGA. These weights can be adjusted by the user for different optimization targets if a lower LUT and FF usage are desirable.

Chimera also has several adjustable hyper-parameters that can be tuned. However, thanks to the robustness of the underlying algorithm, tuning them is unnecessary in most cases. In fact, in the experiments, the same hyper-parameter values are used for all benchmarks in the Rosetta suite. Specifically, the mutation rate of the features in the evolutionary and mutational engine are both set to $0.1$. Recall that the probability evaluation is also used for evaluating the quality of a design point, the $\delta$ value is set to 1.0 for both mutational and evolutionary engine and set to 1.5 for the random proposal engine. Also, for the evolutionary engine, the threshold for selecting the population is set to 1.2. This means that the points whose resource usage is lower than 1.2$\times$ of the latency-wise projected point on the Pareto frontier are included in the population.

\begin{table}[h!]
\vspace{-1mm}
\caption{Comparison between the explored lowest-latency points and the hand-tuned design points}
\centering
\begin{tabular}{|c|c|c|c|c|}
\hline
& \multicolumn{2}{c|}{Hand-tuned} & \multicolumn{2}{c|}{Chimera} \\
\hline
Name     & Latency & Resource & Latency & Resource\\
\hline
3DR      & 5008 & 0.0356  & 5011 \textcolor{Orange}{(+0.06\%)} & 0.0336 \textcolor{ForestGreen}{(-5.6\%)} \\
\hline
BNN      & 409 & 0.0758 & 409 \textcolor{ForestGreen}{(-0.00\%)} & 0.0757 \textcolor{ForestGreen}{(-0.00\%)}\\
\hline
DR       & 18670 & 0.129 & 18790 \textcolor{Orange}{(+0.6\%)} & 0.122 \textcolor{ForestGreen}{(-5.4\%)}\\
\hline
FD       & 54802 & 0.156 & 54290 \textcolor{ForestGreen}{(-0.93\%)} & 0.156 \textcolor{ForestGreen}{(-0.00\%)}\\
\hline
OF       & 4496 & 0.155 & 4496 \textcolor{ForestGreen}{(-0.00\%)} & 0.155 \textcolor{ForestGreen}{(-0.00\%)} \\
\hline
SF      & 11755 & 0.0793 & 11755 \textcolor{ForestGreen}{(-0.00\%)} & 0.0752 \textcolor{ForestGreen}{(-5.17\%)} \\
\hline
\end{tabular}
\vspace{-2mm}
\label{tab:results}
\end{table}

The final results for the benchmarks are presented in Fig. \ref{fig:results}. The red dot represents the hand-tuned result from the benchmark suite, the yellow curve represent the Pareto frontier explored by Chimera in a single exploration run, and the blue curve represents the Pareto frontier explored with only random proposal engine. Table \ref{tab:results} presents the detailed comparison between the hand-tuned design point and the lowest-latency point explored on weighted resource usage and latency. The latency number are measured in $\mu s$ and the resource numbers are the weighted resource usage. The percentage numbers shows the improvement or degradation of the explored design point, and all matching or superior numbers are marked in green.

From Fig. \ref{fig:results} and Table \ref{tab:results}, we can observe that Chimera can discover design points that are equally optimal or superior to the hand-tuned design. When optimizing for low latency, the design points found by Chimera consume fewer resources while having the same or negligibly higher latency. The diagrams also show that, in general, the introduction of EA (both evolutional and mutational engine) help Chimera to discover more optimal Pareto frontiers, especially on finding low-latency or low-resource usage points. This is because, when searching for extreme points, we usually need to combine several beneficial combination of pragmas to find a new point. However, as mentioned before, while the random method is effective on introducing new information to the dataset, it is less directed. It is apparent that generating the correct combination of pragmas purely from random combination can be extremely hard, but the EA methods help combining known good design points near the Pareto frontier, therefore has a higher chance of getting new extreme design points. EA also makes the algorithm more prune to local minima, and within a fixed number of points to explore, as shown in Fig \ref{fig:results} (d), it is possible that the EA algorithms wastes more attempts on finding marginally more optimal points around the current Pareto frontier. But in such a case, if given more time to explore, the Thompson sampling method will help Chimera to switch to random exploration and escape.

It should be noted that, as a multi-objective optimization method, Chimera is not only optimizing for lower latency but also finding a Pareto curve. The designers of  Rosetta only optimized the benchmarks for lower latency, so they miss the better designs with slightly higher latency but significantly lower resource usages. \emph{For example, in the 3D rendering benchmark, the latency of the hand-tuned design is 5008 $\mu s$ and consumes 16512 FFs, whereas the design point at the elbow of the curve (marked with a green dot in Fig. \ref{fig:results_orig_3dr}) with a latency of 5016 $\mu s$ only consumes 12229 FFs, which is a 26\% reduction.} Such a design point can be actually more desirable in practice, since it saves more resources that can potentially be used by other components of a larger design on the FPGA, or reduces congestion in the place and route process, leading to better overall performance.

\begin{minipage}{\columnwidth}
\begin{lstlisting}[language=C, caption={Code snippet from the face detection benchmark}, label={lst:face_detect}]
const int WINDOW_SIZE = 25;
uint18_t _II[WINDOW_SIZE*WINDOW_SIZE];
#pragma HLS array_partition variable=_II complete dim=0
COPY_LOOP1: for (int i = 0; i < WINDOW_SIZE; i ++ ){
    #pragma HLS unroll
    COPY_LOOP2: for (int j = 0; j < WINDOW_SIZE; j ++ ){
        #pragma HLS unroll
        _II[i*25+j] = II[i][j];}}
\end{lstlisting}
\vspace{-1mm}
\end{minipage}    

One of the main advantages of Chimera is that it can search a wider range of directives and find correlations that are not obvious to human designers. Consider the code snippet in Listing \ref{lst:face_detect} taken from the original hand-tuned design of the face detection benchmark. As we can see, when handling the parallel access to the array from unrolled loops, the human designer simply chooses to completely partition the array. However, while this can lead to low latency in general, it is unnecessary to completely partition the array in this case. On the contrary, Chimera found that cyclic partitioning with a factor of 25 can lead to the same latency with lower resource usage. The original intention of the human designer is to completely partition the array \texttt{\_II}, such that the elements are scattered as 625 individual registers. This will allow simultaneous access to each of the elements, so all the accesses from the unrolled loops can be satisfied at the same time. However, in the synthesized hardware, although both loop levels are completely unrolled, it is likely that not all 625 memory accesses are scheduled in the same cycle due to other internal optimizations or limitations of HLS. As a result, the data access bandwidth of the fully partitioned array is underutilized. This result shows that the human designer cannot fully discover the complicated interaction inside the HLS to determine the precise number of parallel accesses to the array to select the minimum partitioning factor. So, in this case, the hardware resource is wasted on the partitioning.

In terms of the total time taken for the DSE, the exploration took less than 24 hours for each of the benchmarks on the test machine. Such a result demonstrates the practicality of Chimera in real-world scenarios. For example, the designer can leave the exploration running on a server overnight and the optimized results can be available the next day. This improvement can be largely attributed to the error and timeout prediction models added, such that it does not need to wait for a timeout for many of the un-synthesizable design points.

\section{CONCLUSION}
In conclusion, Chimera is a novel ML-driven software tool that facilitates the DSE process for tuning HLS directives. It combines the strengths of multiple optimization algorithms to form an hybrid efficient DSE method. Which helps reducing the  The simpler interface also means that it can be easily adopted by new users of HLS tools.

For future improvements on Chimera, one possibilities is to use an adaptive mutation rate of the evolutionary and mutational engines. By using random forest models, it is possible to analyze the importance of each input feature and adjust the mutation rate accordingly to further improve the quality of the proposed point.

It also is worth noting that the core of Chimera can be ported to other HLS tools or other black-box optimization problems. For instance, it can potentially be applied to problems such as automated architecture exploration for ML accelerators \cite{yazdanbakhsh2021apollo}. Therefore, we will also release the source code to the public to enable future collaborations on Chimera.

\section{ACKNOWLEDGEMENTS}
This work is published in the 22nd International Conference on Intelligent Data Engineering and Automated Learning (IDEAL 2021) and is supported in part by the Xilinx Center of Excellence and Xilinx Adaptive Compute Clusters (XACC) program at the University of Illinois Urbana-Champaign.

\bibliographystyle{IEEEtran}

{\small
\bibliography{IEEEabrv,thesisrefs}
}

\end{document}